\documentstyle[twocolumn,aps,prl]{revtex}

\renewcommand{\vec}[1]{{\bf #1}}
\newcommand{\be}{\begin{equation}}
\newcommand{\ee}{\end{equation}}

\def\draftfig{0}

\ifnum\draftfig=0
  \input epsf.sty
\fi

\begin{document}

\draft

\preprint{}

\twocolumn[\hsize\textwidth\columnwidth\hsize\csname %
@twocolumnfalse\endcsname

\title{Macroscopic Quantum Coherence in Molecular Magnets}

\author{Alain Chiolero and Daniel Loss}

\address{Department of Physics and Astronomy, University of Basel,\\
Klingelbergstrasse 82, 4056 Basel, Switzerland}

\date{\today}

\maketitle

\begin{abstract}

We study macroscopic quantum coherence in antiferromagnetic molecular
magnets in the presence of magnetic fields. Such fields generate
artificial tunnel barriers with externally tunable strength.  We give
detailed semi-classical predictions for the tunnel splitting in
various regimes for low and high magnetic fields. We show that the
tunneling dynamics of the N\'eel vector can be directly measured via
the static magnetization and the specific heat.  We also report on a
new quantum phase arising from fluctuations. The analytic results are
complemented by numerical simulations.

\end{abstract}


]

\narrowtext

Quantum spin dynamics in mesoscopic magnets has received much
attention over the recent years, both from experiment and from theory
\cite{NATO94}. A number of nanosized particles in the
superparamagnetic regime have been identified as promising candidates
for the observation of macroscopic quantum phenomena (MQP) such as the
tunneling of the magnetization out of a metastable potential minimum,
or, more strikingly, macroscopic quantum coherence (MQC), where the
magnetization (or the N\'eel vector) tunnels coherently between
classically degenerate directions over many periods.  On one hand,
these phenomena are interesting from a fundamental point of view as
they extend our understanding of the transition from quantum to
classical behavior.  On the other hand, the measurement of MQP
quantities such as the tunnel splitting provides independent
information about microscopic parameters such as anisotropies and
exchange constants.

A prominent example of such MQC behavior that has attracted wide
attention (but also scrutiny) is the antiferromagnetic ferritin
\cite{Ferritin}.  More recently, molecular magnets \cite{Gatt94} such
as the ferric wheel or Mn$_{12}$ have emerged as promising candidates
for the experimental observation of MQP \cite{Fried96,Barb96} mainly
for three reasons.  First, molecular magnets have well-defined
structures and magnetic properties.  Thus, precise values for the
tunneling rates can be calculated.  Second, molecular magnets can be
produced as single crystals that contain a macroscopic number of
identical magnetic subunits, which leads to a natural amplification of
the single-unit signal. Third, the typically high symmetry of these
magnets reduces the number of independent parameters.

In this letter we discuss novel tunneling scenarios in
antiferromagnetic (AFM) molecular magnets. A key feature of our
discussion is to exploit the well-known fact \cite{RefField} that an
effective anisotropy can be generated in an AFM by applying a magnetic
field. Thus it is possible to create tunnel barriers that are tunable
by an external parameter.  Evidently, such control parameters are
highly desirable as they open the door to systematic tests of MQC.  We
concentrate on ring-like structures such as ${\rm Fe}_6$, ${\rm
Fe}_{10}$, ${\rm V}_8$ \cite{Gatt94}, where the spins interact with
their nearest neighbors via exchange coupling.  In particular, we show
that the tunneling rates become field dependent and thus can be
measured via the static magnetization and (less surprisingly) also via
the Schottky anomaly of the specific heat.  Our discussion is based on
the non-linear sigma model (NLsM) that includes anisotropies and
magnetic fields.  The quantum dynamics of the N\'eel vector is then
studied by instanton methods. Such methods are semiclassical in
nature, i.e.\ valid for large spins and in the continuum limit.  To
cover the small end of the size scale, we performed ab initio
numerical calculations; they agree well with the analytic results in
the regime where a comparison is possible.  We find several distinct
tunneling regimes, depending on the ratio of crystalline anisotropy to
magnetic field.  Motivated by recent measurements on single-crystal
Fe$_{10}$ which indicate the presence of an anisotropy axis \cite
{Harris}, we give estimates of these MQC corrections in the
magnetization and the specific heat, and we show that they are within
experimental reach.

We consider a ring-like molecular magnet, modeled as $N$ spins $s$
regularly spaced on a circle lying in the $xy$-plane, with $N$ even.
The Hamiltonian is ($\vec{S}_{N+1}\equiv\vec{S}_1$)
\be
H=J\sum_i\vec{S}_i\cdot\vec{S}_{i+1}+\sum_i U_i(\vec{S}_i)+
\hbar\vec{h}\cdot \sum_i\vec{S}_i{,}
\label{model}
\ee
with AFM exchange coupling $J>0$, and where $U_i(\vec{S}_i)$ is the
crystalline anisotropy at site $i$, $\vec{h}=\gamma\vec{B}$, with
$\vec{B}$ being the magnetic field, $\gamma=g\mu_B/\hbar$, and $g$ is
the electronic g-factor. For simplicity, we assume that both $g$ and
$J$ are isotropic, and that the point symmetry of the molecule is that
of a ring.  Up to second-order in the spin variables, the most general
form of the anisotropies is then $U_i(\vec{S}_i)=
\tilde{k}_zS_{i,z}^2+\tilde{k}_r(\vec{S}_i\cdot\vec{e}_i)^2$, where
$\tilde k_z$ and $\tilde k_r$ are the axial and radial anisotropies,
respectively, and $\vec{e}_i$ is a unit vector at site $i$ pointing
radially outwards.  We also assume that $J\gg \tilde k_z$, $\tilde
k_r$, which is typically the case.

We now derive an effective Lagrangian (NLsM) describing the low-energy
physics of (\ref{model}) by extending standard techniques
\cite{Frad91} to include magnetic fields. We introduce spin
coherent-states, and decompose the local spin as
$\vec{S}_i=(-1)^is\vec{n}+\vec{l}_i$, where the N\'eel vector
$\vec{n}$ (with $\vec{n}^2=1$) is taken as uniform for our small
system, and $\vec{l}_i$ is the fluctuation at site $i$ (with
$\vec{l}_i\cdot \vec{n}=0$). After integrating out the
$\vec{l}_i$ \cite{constraint}, and keeping only the lowest-order terms
in $\tilde k_z/J$ and $\tilde k_r/J$, we obtain the Euclidean
Lagrangian
\be
L_E={N\hbar^2\over8J}\left[-(i\vec{n}\wedge\dot{\vec{n}}-\vec{h})^2+
(\vec{h}\cdot\vec{n})^2\right]+
N{k}_zs^2n_z^2{,}
\label{Lagrang}
\ee
with a single effective axial anisotropy ${k}_z=\tilde k_z-\tilde
k_r/2$.  Note that the magnetic field has two effects. First, it
creates a hard axis anisotropy along its direction.  This was already
noticed in the context of spin chains \cite{RefField}, and is easy to
interpret. The spins can gain Zeeman energy by canting towards the
magnetic field, and this effect is maximal when the N\'eel vector is
perpendicular to the field. What makes this anisotropy interesting for
our purposes is that it is tunable by an external field.  Second, a
phase factor arises from the cross term
$2i\vec{h}\cdot(\vec{n}\wedge\dot{\vec{n}})$.  This is a crucial
difference to the results of Ref.\ \onlinecite{RefField}, which will
have important consequences at high fields.

Depending on the sign of ${k}_z$ and on the orientation of the field,
various scenarios can be envisaged \cite{Chiolero2}. For lack of space
we restrict ourselves to the most interesting case where the field is
in the ring plane, $\vec{B}=(B_x,0,0)$, and perpendicular to a hard
axis, i.e.\ ${k}_z>0$.  The potential energy has then two minima at
$\vec{n}=\pm\vec{e}_y$. Tunneling of the N\'eel vector between these
classically degenerate states results in a tunnel splitting of the
ground state energy, which can be calculated using instanton methods
\cite{Raja}.  For the semi-classical dynamics two kinds of hard axis
anisotropies compete, the crystalline one, $N{k}_zs^2$, and that
induced by the field, $N(g\mu_BB_x)^2/8J$. Let us introduce their
ratio $\lambda=8s^2J{k}_z/(g\mu_BB_x)^2$.  For low fields
($\lambda\gg1$), the hard-axis is the $z$-axis, and the N\'eel vector,
staying close to the $xy$-plane, tunnels via the $x$-axis. For high
fields ($\lambda\ll1$) on the other hand, the hard-axis is the
$x$-axis, and the N\'eel vector tunnels via the $z$-axis while staying
in the $yz$-plane.  Without the phase term
$2i\vec{h}\cdot(\vec{n}\wedge\dot{\vec{n}})$ in (\ref{Lagrang}), the
crossover would occur for $\lambda=1$, i.e.\ for a field
$B_x=s\sqrt{8J{k}_z}/g\mu_B$. As we will see, this extra term reduces
the critical field.

We first concentrate on the high-field regime ($\lambda\ll1$). In this
case, the N\'eel vector is conveniently parameterized as ${\bf n} =
(\cos\theta,\sin\theta\sin\phi,\sin\theta\cos\phi)$.  We then find
that the instanton solution minimizing the action belonging to
(\ref{Lagrang}) moves in the $yz$-plane \cite{Lorentz} with the
frequency $\omega_{\rm hf}=s\sqrt{8J{k}_z}/\hbar$, and action
\be
S_{\rm hf}/\hbar=Ns\sqrt{2{k}_z/J}\pm i\pi{Ng\mu_BB_x\over4J}{,}
\label{actionHF}
\ee
where the upper and lower signs correspond to instantons and
anti-instantons, respectively.  The phase term in Eq.\
(\ref{actionHF}) arises because the spins cant towards the field,
thereby acquiring an additional geometric phase factor.

For the calculation of the fluctuation determinants, it is convenient
to pass to dimensionless variables in (\ref{Lagrang}).
We write the action as $S=\hbar
N{g\mu_BB_x/8J}\int\!\!  d\bar{\tau}\,(\tilde{L}-h_x)$, with
\be
\hbar\omega_{\rm lf}=g\mu_BB_x{,}\quad
\tilde{L}=\dot{\vec{n}}^2+2i(\vec{n}\wedge\dot{\vec{n}})_x+n_x^2+
\lambda n_z^2{,}
\label{Lagrang_red}
\ee
where time is rescaled as $\bar{\tau}=\omega_{\rm lf}\tau$.
Expanding around the instanton solution up to second-order, one sees
that the $\theta$- and $\phi$-fluctuations are decoupled. For the
$\theta$-fluctuation determinant we find in leading order $\exp\{\pm i
{\pi\over 2}+{\cal O}(\lambda^{3/2})\}$.  This is a new quantum phase
distinguishing instantons from anti-instantons. Its occurrence is
surprising, because it is due to quantum {\em fluctuations}, and not
due to a topological term in the action, in marked contrast to phases
arising usually in spin problems \cite{Frad91,Loss92}. Moreover, this
new quantum phase does {\em not} depend on the spin $s$.  The
fluctuation determinant for $\phi$ is standard \cite{Raja}, and by
summing over all configurations \cite{Loss92}, we finally find for the
tunnel splitting of the ground state in the high-field regime,
\begin{eqnarray}
\Delta_{\rm hf}&=&8\hbar\omega_{\rm hf}
\sqrt{{\rm Re}S_{\rm hf}\over2\pi\hbar}
e^{-{\rm Re}S_{\rm hf}/\hbar}
\left|\sin\left({\pi\over2}{Ng\mu_BB_x\over2J}\right)\right|{.}
\label{dEhf}
\end{eqnarray}
Note that the tunnel splitting oscillates with the B-field as a result
of interference between quantum spin phases
(the new additional phase induces a shift from the usual cosine
\cite{Loss92} to a sine).
Naively, one would have expected no field dependence, since
both the tunneling barrier and the attempt frequency are constant.
All these features are nicely confirmed by independent numerics (see
below and Fig.\ 2).

Next, we consider the low-field regime ($\lambda\gg1$), where we use
the parametrization ${\bf n} = (\sin\theta\cos\phi,
\allowbreak\sin\theta\sin\phi, \allowbreak\cos\theta)$.  We start by
integrating out the $\theta$-fluctuations to obtain an effective
action. The term $(\vec{n}\wedge\dot{\vec{n}})_x$ in the Lagrangian
(\ref{Lagrang_red}) forces the instantons out of the
$xy$-plane. However, for $\lambda\gg1$ the deviations are small. Thus,
we write $\theta=\pi/2+\vartheta$, and expand the Lagrangian to
second-order in $\vartheta$. This gives
$\tilde{L}\approx\dot{\phi}^2+\cos^2\phi+
4i\vartheta\dot{\phi}\cos\phi+\vartheta G\vartheta$, where
$G=-\partial_{\bar{\tau}}^2+\lambda-\cos^2\phi-\dot{\phi}^2$.
Integrating out $\vartheta$ is now straightforward.  For $\lambda\gg1$
we find \cite{kernel} that $G\approx\lambda$.  Hence, we end up with
the effective Lagrangian
\be
\tilde{L}_{\rm eff}=
\Bigl(1+{4\over\lambda}\cos^2\phi\Bigr)\dot{\phi}^2+\cos^2\phi{.}
\label{Ltilde}
\ee
A simple quadrature shows that the instantons of this Lagrangian have
an action $\tilde{S}_{\rm eff}=4(1+4/3\lambda+{\cal
O}(\lambda^{-2}))$.  Reinstating full units, we finally get for the
action in the low-field regime (neglecting corrections of order
$\lambda^{-2}$)
\be
S_{\rm lf}/\hbar=N{g\mu_BB_x\over2J}
\left(1+{1\over6}{(g\mu_BB_x)^2\over s^2J{k}_z}\right){.}
\ee
Comparison with Eq.\ (\ref{actionHF}) shows that the crossover between
the low- and high-field regimes occurs for a magnetic field
$B_x=\alpha s\sqrt{8J{k}_z}/g\mu_B$, with
$\alpha=[(3+\sqrt{10})^{1/3}-(3+\sqrt{10})^{-1/3}]/2\approx0.64$, a
sizeable reduction over the result that would follow from neglecting
the phase term $2i\vec{h}\cdot(\vec{n}\wedge\dot{\vec{n}})$ in
(\ref{Lagrang}).

Next we determine the fluctuation-determinant. This raises one
problem. While the action for the instanton solutions of
(\ref{Ltilde}) is easy to calculate, the solutions themselves cannot
be obtained analytically. However, being only interested in the regime
$\lambda\gg1$, we can approximate this determinant by the one obtained
for the fluctuations around the instantons of the Lagrangian
$\bar{L}=\dot{\phi}^2+\cos^2\phi$. This results in
\be
\Delta_{\rm lf}=8\hbar\omega_{\rm lf}\sqrt{S_{\rm lf}\over2\pi\hbar}
e^{-S_{\rm lf}/\hbar}{.}
\ee
The low-field splitting decreases (roughly) exponentially with the
field. This is easily interpreted: The tunneling barrier increases
quadratically with the field, whereas the attempt frequency increases
linearly.

\begin{figure}

\ifnum\draftfig=1
  \vspace*{0cm}
\else
$$\epsffile{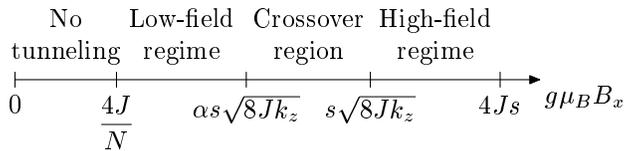}$$
\fi

\caption{Summary of the various regimes we have obtained as functions of
the applied magnetic field ($\alpha\approx0.64$).}
\label{regimes}

\end{figure}

We complete our derivation by discussing the range of validity of our
results. A necessary condition to have tunneling (in the ground state)
is that the barrier $\Delta U$ be much larger than half the attempt
frequency $\omega$\cite{Inst_crit}.  Application of this criterion to
the low- and high-field regimes yields two conditions,
$g\mu_BB_x\gg4J/N$, and $Ns\sqrt{{k}_z/2J}\gg1$.  The effective
Lagrangian (\ref{Lagrang}) was derived under the assumption of (local)
N\'eel order. Hence, the Zeeman energy must be smaller than the
exchange energy, i.e.\ $g\mu_BB_x\ll 4Js$.
Finally, our expression for high-field tunnel splitting is valid for
$\lambda\ll1$, i.e.\ for $g\mu_BB_x\gg s\sqrt{8J{k}_z}$, whereas the
low-field predictions hold if $\lambda\gg1$, i.e.\ for $g\mu_BB_x\ll
s\sqrt{8J{k}_z}$.  We summarize in Fig.\ \ref{regimes} the various
regimes and critical fields we have obtained.

We now turn to the question of how to observe the tunnel splitting.
In contrast to previous cases \cite{Barb90} such as ferritin
\cite{Ferritin} it is not possible to observe the switching of the
N\'eel vector via an excess spin, since even if such an excess moment
were present, it is easily seen \cite{Chiolero2} that it would always
point along the magnetic field and not along the N\'eel vector.
However, the dynamics of the N\'eel vector could be observed via
resonances (occurring at $\Delta$) in the NMR spectrum, which provides
local spin information.
An entirely different approach, which is only possible because the
tunnel splitting is B-field dependent, is to measure the static
magnetization $\vec{M}=-\bigl\langle
g\mu_B\sum_i\vec{S}_i\bigr\rangle$ as a function of applied field.
Indeed, we have seen that the two lowest energies are tunnel split by
$\Delta$ and are separated from the other levels by an energy
$\hbar\omega\gg \Delta$, where $\Delta$ and $\omega$ are approximated
by $\Delta_{\rm lf}$, $\Delta_{\rm hf}$, and $\omega_{\rm lf}$,
$\omega_{\rm hf}$, respectively, depending on the field. At low
temperatures, such that $k_BT\ll\hbar\omega$, the magnetization along
the $x$-axis is then found to be
\be
M_x=\left({N\over8J}g\mu_BB_x-{1\over2}\right)g\mu_B+
{\Delta'\over2}
\tanh\left(\Delta\over2k_BT\right){.}
\label{Mx}
\ee
Note that the first two terms in Eq.\ (\ref{Mx}) only give a linear
dependence on $B_x$. The last term shows that deviations are
proportional to $\Delta'=\partial\Delta/\partial B_x$.

The Lagrangian (\ref{Lagrang}) is not limited to the tunneling regime,
but also covers the nearly-free limit of small ${k}_z$, such that
$Ns\sqrt{{k}_z/2J}\ll1$. This regime is most conveniently
studied\cite{Steps} in terms of the corresponding Hamiltonian which is
of rigid rotor type, $H_{\rm rot}={2J\over N \hbar^2}\vec{L}^2+
\gamma\vec{L}\cdot\vec{B}+N{k}_zs^2n_z^2$,
where $\vec{n}$ and the angular momentum $\vec{L}$
satisfy standard commutation relations
$[L_j,n_k]=i\hbar\epsilon_{jkl}n_l$. For ${k}_z=0$, the ground-state
is the state $|l,l\rangle$, with $l=\lfloor
g\mu_BB_xN/4J\rfloor$. Hence, the magnetization $M_x=lg\mu_B$ consists
of steps of height $g\mu_B$, occurring with a period $g\mu_B\Delta
B_x=4J/N$ \cite{period}. This agrees with previous results obtained in
the absence of anisotropies and tunneling \cite{Gatt94}.  A small
value of ${k}_z$ (before tunneling sets in) leads to a rounding of the
steps, as is easily seen from perturbation theory.  For larger $k_z$
such that $\Delta \ll \hbar\omega$ the N\'eel vector no longer freely
rotates but becomes strongly localized along the easy axis. As a
consequence, the steps in the magnetization vanish, see
(\ref{Mx}). Conversely, notice that whatever the value of $k_z\ll J$
sharp steps are recovered if the magnetic field is applied {\em along}
the hard-axis. We have also confirmed this picture by direct numerical
diagonalization of $H_{\rm rot}$.

The tunnel splitting is also reflected in other thermodynamic
quantities. For example, the low temperature specific heat exhibits a
characteristic Schottky anomaly \cite{Pathria}, $c_V=
k_B(\Delta/2k_BT)^2{\rm sech}^2(\Delta/2k_BT)$, with a peak of height
$0.64\, k_B$ at a temperature $T\approx0.6\Delta/k_B$. The location of
this peak thus gives the tunnel splitting.

\begin{figure}

\ifnum\draftfig=1
  \vspace*{0cm}
\else
\begin{center}
\epsfxsize=8cm
$$\epsffile{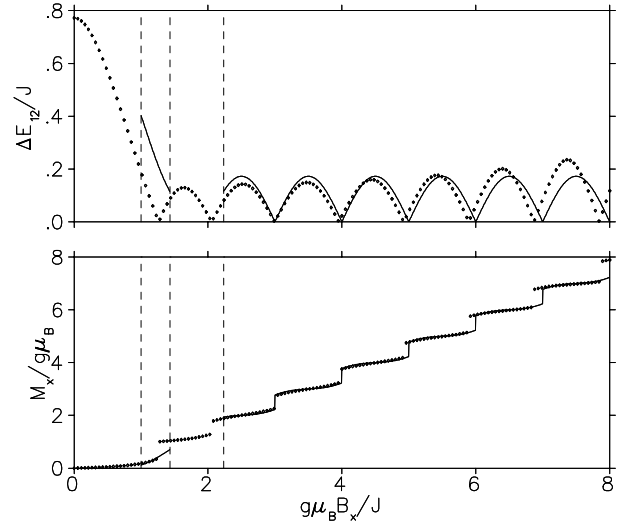}$$
\end{center}
\fi

\caption{Results for the energy splitting $\Delta E_{12}$ between
the lowest two states (upper part) and the magnetization $M_x$ at
$T=0$ (lower part) for a system of $N=4$ spins $s=5/2$ with $k_z=J/10$
and $k_r=0$. The symbols give the results of a direct diagonalization
of the Hamiltonian (\protect\ref{model}), the continuous lines give
the semi-classical predictions. The dashed vertical lines indicate the
various critical fields (see Fig.\ \protect\ref{regimes}).}
\label{numerics}

\end{figure}

The semi-classical analysis presented so far applies strictly speaking
only to a sizable number of spins with $s\gg 1$.  However, as is often
the case with such methods the results are valid (at least
qualitatively) even down to a few spins of small size.  This
expectation is indeed confirmed by direct numerical simulations which
we have performed on Hamiltonian (\ref{model}). Results for $N=4$ and
$s=5/2$, and for some typical values  $\tilde{k}_z=J/10$,
and $\tilde k_r=0$, are presented in Fig.\ \ref{numerics}.  We note
that since most symmetries are broken in (\ref{model}), larger system
sizes become quickly inaccessible to numerics.  The agreement with the
semi-classical prediction is satisfactory in the high-field regime.
Since for our test system the low-field regime becomes vanishingly
small, i.e.\  $1\ll g\mu_BB_x/J\ll\sqrt{5}$, we cannot expect to find
good quantitative agreement in this regime. Still, we can see from
Fig.\ \ref{numerics} that at the qualitative level the numerical and
semiclassical approach show reasonable agreement.  We have also
calculated numerically the matrix elements of the staggered
magnetization and could confirm the tunneling picture. Thus, our
theoretical predictions give reasonably good results even for a very
small cluster (and similarly for rings with larger $N$ but smaller $s$
\cite{Chiolero2}). Obviously, the accuracy of the semiclassical
results will improve for larger systems.

To support the experimental relevance of our results, we give some
estimates for the ferric wheel, Fe$_{10}$, for which $N=10$, $s=5/2$,
$J/g\mu_B=10\,{\rm T}$ \cite{Gatt94}.
While magnetization measurements have been reported
\cite{Gatt94}, no conclusive comparison with our theory is
possible presently since they have been performed on polycrystalline
samples with random orientation of the anisotropy axis, whereas the
tunneling effect discussed here requires the $B$-field to have a fixed
orientation with respect to such an axis.
However, from the well-defined steps that have been observed and from
recent single-crystal measurements \cite{Harris} one can infer that
the magnitude of $k_z/J$ is small, and of the order of $0.03$
\cite{Harris}.  The low-field regime extends then from $4\,{\rm T}$ to
$7.8\,{\rm T}$, with a tunnel splitting $\Delta_{\rm lf}/h$ decaying
exponentially from (roughly) $6\cdot10^{10}\,{\rm Hz}$ to
$4\cdot10^9\,{\rm Hz}$. Correspondingly, the Schottky peak of the
specific heat shifts from $1.6$ to $0.12\,{\rm K}$, while the
tunneling corrections in the magnetization range from $60\,\%$ down to
$16\%$ of $\mu_B$.  The high-field regime starts at $12\,{\rm T}$,
with the tunnel splitting having oscillations of magnitude
$\Delta_{\rm hf}/h\approx 6\cdot 10^9\,{\rm Hz}$ and period $4\,{\rm
T}$.  The tunneling corrections in the magnetization reach at their
peak $17\,\%$ of $\mu_B$. The Schottky peak oscillates between zero
and $0.2\,{\rm K}$. The crossover temperature to the quantum regime,
$T_c=\hbar\omega/4k_B$, is in the $1$--$4\,{\rm K}$ range. Finally we
remark that the same numbers apply to the high-field regime of an
easy-axis system \cite{Chiolero2}. Hence, all quantities appear to be
well within experimental reach.

We are grateful to D.D.\ Awschalom and J.\ Harris for useful
discussions and for providing us with unpublished data.
One of us (DL) is grateful for the hospitality of the
ITP Santa Barbara (US NSF Grant No.\ PHY94-07194), and of the INT
Seattle, where part of this work has been performed.

\begin{references}
\bibitem{constraint} Note that due to the magnetic field
special care must be given to the constraint
$\vec{n}\cdot\vec{l}_i=0$.
\bibitem{Lorentz} Since the magnetic force is proportional to
$\vec{h}\wedge\dot{\vec{n}}$, it is parallel to $\vec{n}$ for any
trajectory in the $yz$-plane and has therefore no effect.
\bibitem{kernel} In the dimensionless units used here, the
tunneling time is of order $1$. Hence, $\partial_{\bar{\tau}}^2
\sim{\cal O}(1)$, $\dot{\phi}\sim{\cal O}(1)$, and obviously
$\cos\phi\sim{\cal O}(1)$; all these terms are therefore much smaller
than $\lambda$ in the small-field regime.
\bibitem{Inst_crit} In both the low- and high-field regimes the action
is approximately equal to $4\Delta U/\omega$. Hence, our criterion
also implies that the action is larger than $2\hbar$, which is also
necessary for the dilute-instanton approximation to hold.
\bibitem{Steps} A complementary analysis can be given directly from the
Lagrangian, but the treatment gets more involved \protect\cite{Chiolero2}.
\bibitem{period} Note that this period coincides with the one obtained
for the oscillations in $\Delta_{\rm hf}$, Eq.\ (\ref{dEhf}).
\end{references}
\end{document}